\documentclass[12pt]{article}

\usepackage{bbold}
\usepackage{amssymb}
\usepackage{slashed}
\usepackage{graphics}

\declareslashed{}{/}{.1}{0}{E}

\def\r{\rho}
\def\s{\sigma}

\def\Sl#1{\slashed{#1}}

\def\d{\partial}
\def\m{\mu}
\def\n{\nu}

\def\be{\begin{equation}}
\def\ee{\end{equation}}
\def\beq{\begin{equation}}
\def\eeq{\end{equation}}
\def\bea{\begin{eqnarray}}
\def\eea{\end{eqnarray}} 
\def\beqa{\begin{equation}\begin{array}{l}}
\def\eeqa{\end{array}\end{equation}}

\def\eqn#1{(\ref{#1})}

\def\eqref#1{eq.~(\ref{eq:#1})}

  \def\g{\gamma}

\def\L{{\it\Lambda}}

\def\w{\omega}

\newcommand{\ul}[1]{\hspace{.5mm} \underline{#1}}

\begin{document}

\thispagestyle{empty}
\begin{flushright}
\framebox{\small BRX-TH~491
}\\
\end{flushright}

\vspace{.8cm}
\setcounter{footnote}{0}
\begin{center}
{\Large{\bf 
Null Propagation of Partially Massless Higher Spins in (A)dS and Cosmological
Constant Speculations}
    }\\[10mm]

{\sc S. Deser
and A. Waldron
\\[6mm]}

{\em\small  
Physics Department, Brandeis University, Waltham,
MA 02454, 
USA\\ {\tt deser,wally@brandeis.edu}}\\[5mm]

\bigskip

\bigskip

{\sc Abstract}\\
\end{center}

{\small
\begin{quote}

We show explicitly that all partially and strictly massless 
fields with spins $s\leq3$ in (A)dS have null propagation.
Assuming that this property 
holds also for $s>3$, we derive the mass-cosmological constant tunings 
required to yield all higher spin partially massless theories.
As $s$ increases, the unitarily allowed region for massive spins
is squeezed around $\L=0$, so that an infinite 
tower of massive particles forces vanishing $\Lambda$.  We 
also speculate
on the relevance of this result to string theory and supergravity
in (A)dS backgrounds.

\bigskip

\bigskip

\end{quote}
}

\newpage





\section{Introduction}

The cosmological constant ``problem'' has seen many incarnations.
Current observational evidence points to a 
small but positive, de Sitter (dS), 
value for $\L$~\cite{Carroll:2000fy}, while recent 
theoretical advances relating gauge and gravitational
theories are based on strings
in $\L<0$, Anti de Sitter (AdS), backgrounds~\cite{Maldacena:1998re}, 
the sign also required by 
conventional cosmological supergravity~\cite{Pilch:1985aw}.
In the context of $\L\neq0$ spaces, a recent 
analysis 
of free, higher ($s>1$) spin fields has led to 
dramatic results~\cite{Deser:2001pe}:  
In addition to the usual Minkowski possibilities of
$2$ strictly massless or $2s+1$ massive
propagating degrees of freedom (DoF), there are intermediate 
partially massless theories corresponding to new gauge
invariances for particular tunings of the mass $m^2$ to
$\Lambda$ that remove lower helicity states from the 
spectrum. These novel theories divide the $(m^2,\L)$
plane into unitarily allowed and forbidden regions.
While explicit constructions were provided in~\cite{Deser:2001pe}
for all $s\leq3$, two important questions were left open:
\begin{enumerate}
\item Do these new partially massless theories propagate on the null
cone?
\item What $m^2$~:~$\Lambda$ tunings are necessary to achieve partial
masslessness for arbitrary $s>3$?
\end{enumerate}
In this Letter, we answer both questions and draw some possible
consequences: Null propagation is 
verified for all the $s\leq3$ gauge theories
and assuming that this result persists for
$s>3$ provides a simple derivation of the higher spin
$m^2$~:~$\Lambda$ tunings.
Physically, this assumption just extends the usual flat space wisdom
that gauge invariance implies lightlike propagation, but there
is a second compelling reason to assert its validity: The existence
of partially massless theories is related directly to 
representations of the (A)dS  spacetime algebra.
A study of the $s=2$ partially massless 
theory~\cite{Deser:1983tm,Higuchi:1987py} 
prompted a verification 
that (bosonic) representations with missing lower helicities
were indeed possible~\cite{Higuchi:1987wu}. Our (local) 
models realize precisely these
representations and construct their fermionic counterparts.

A potentially wide-ranging consequence of our results is that
as $s$ increases, unitary massive
theories can only occupy a region in the $(m^2,\Lambda)$ plane
that is 
squeezed onto the line $\L=0$. 
[Individual partially massless theories 
are unitary whenever their gauge invariances remove all negative norm states.]
Some ramifications of this result for the cosmological constant problem 
and string theory in constant curvature backgrounds 
are also discussed.

\section{Null Propagation of Partially Massless Bosons}

\label{bose}

Our method is direct: We solve the helicity $\pm s$ field equations  
for all higher spins in a dS
background\footnote{The Riemann tensor in constant curvature spaces is
$
R_{\m\n\r\s}=-\frac{2\L}{3}\,g_{\m[\r}g_{\s]\n}\ .
$
The actions of commutators of covariant
derivatives are summarized by the vector-spinor example
$
[D_\m,D_\n]\,\psi_\r=\frac{2\L}{3}\,g_{\r[\m}\psi_{\n]}
+\frac{\L}{6}\,\g_{\m\n}\psi_\r\, .
$  
Our
metric is ``mostly plus'' and
Dirac matrices are ``mostly hermitean''.}
$$
ds^2=-dt^2+f^2(t)\,d\vec x\,^2\, ,
$$
\be 
f(t)\equiv e^{Mt}\, , \qquad M^2\equiv\L/3\, .
\label{dS}
\ee 
When they are present in the spectrum, all lower helicities
propagate in exactly the same manner (this has been worked
out in great detail~\cite{Deser:2001pe} for $s=2$), so we need not
treat them separately.
In the frame~\eqn{dS}, spatial slices are flat, which allows 
the usual definition of helicity. 
Solutions to the field equations are of Bessel type, and take the form
\be
\Big(\mbox{slowly varying}\Big)\,\times\, 
\exp(i\w u+i\vec k\cdot\vec x)\, ,\qquad \w^2=\vec k\,^2\, ,
\label{wave}
\ee 
whenever
the index $\nu$ of the Bessel function is half integer. 
Here 
\be
u\equiv -\frac{f^{-1}(t)}{M}
\ee
is the conformal time coordinate in terms of which 
\be
ds^2=\frac{1}{M^2u^2}\,\Big(-du^2+d\vec x\,^2\Big)\, .
\ee
Therefore
we can read off the theories propagating on the null cone directly from
the index $\nu$. We will find null propagation 
for all the $s\leq3$ partially massless theories
presented in~\cite{Deser:2001pe}. 

The onshell conditions for a massive spin $s$ field in (A)dS
are
\be
\Big(D^2-m^2-(2+2s-s^2)\,M^2\Big)\,\phi_{\m_1\ldots\m_s}=0\, , \qquad
D.\phi_{\m_2\ldots\m_s}=0=\phi^\r{}_{\r\m_3\ldots\m_s}\, .
\label{bshell}
\ee 
The parameter $m^2$ has been chosen so that $m^2=0$
corresponds to the strictly massless (i.e., helicity $\pm s$ only)
theory. [For $s=0$,
equation~\eqn{bshell} describes a conformally improved scalar.]

The traceless-transverse part of a spatial tensor is 
denoted by the superscript $tt$. 
We project out helicity $\pm s$ by computing only the tt part
of~\eqn{bshell}, which in the frame~\eqn{dS} reads,
\be
\Big(-\frac{d^2}{dt^2}+ (2s-3)\, M\,\frac{d}{dt}
+f^{-2}\, \vec \d\,^2 
- m^2 +2(s-1)\,M^2
\Big)\,\phi^{tt}_{i_1\ldots i_s}=0\, .
\ee  
Fourier transforming $\vec \d\rightarrow i\vec k$, changing coordinates
\be
z\equiv-\frac{kf^{-1}(t)}{M}=ku\, ,\qquad (k\equiv |\vec k|)
\label{change}
\ee
and the field redefinition (suppressing indices)
\be
\phi^{tt}\equiv z^{3/2-s} q\, ,
\ee
yield Bessel's equation
\be
\frac{d^2q}{dz^2}+\frac{1}{z}\,\frac{dq}{dz}
+\Big(1-\frac{\nu^2}{z^2}\Big)\,q=0\, ,
\label{Bessel}
\ee
with index
\be
\nu^2=\frac{1}{4}+s(s-1)-\frac{m^2}{M^2}\, .
\ee
We may now read off the null propagating theories.

\subsubsection*{Examples}

\begin{enumerate}
\item {\em Conformal Scalar}: At $s=0=m$ we obtain $\nu=1/2$
and $q(z)=z^{-1/2}\,\exp(iz)$ which implies
a solution of the form~\eqn{wave}.
The value $\nu=1/2$ also characterizes all the higher spin ``conformal''
theories.
\item {\em Maxwell}: In $d=4$ the $s=1$, $m=0$ vector theory is conformal
and here $\nu=1/2$.
\item {\em Spin 2}: Spin~2 can be either strictly massless at $m=0$ or 
partially massless when $m^2=2M^2$. The latter model, with its 
accompanying scalar gauge invariance, takes the conformal value $\nu=1/2$.
Of course, the $m^2=0$ linearized cosmological Einstein theory also
propagates on the cone, but this is achieved by the solution $\nu=3/2$
for which $q(z)=(z+i)\,z^{-3/2}\,\exp(iz)$.
\item {\em Spin 3}: Here there are 3 possibilities; the strictly massless 
theory at $m=0$ for which $\nu=5/2$ 
($\Rightarrow q(z)=(z^2+3iz-3)\,z^{-5/2}\,\exp(iz)$), 
a partially massless one with helicities $(\pm3,\pm2)$ at
$m^2=4M^2$ with $\nu=3/2$ and finally, a theory with a scalar gauge 
invariance and helicities $(\pm3,\pm2,\pm1)$ with $m^2=6M^2$ and the 
conformal value $\nu=1/2$. Clearly all these theories have null propagation.
\end{enumerate}
Noting that the value of $\nu$ can be associated with the type of 
gauge invariance (for example, the conformal value $\nu=1/2$ always belongs 
to the scalar invariance), we are led to the following conjecture:

\vspace{.4cm}

\noindent {\bf Conjecture}: All partially massless
higher spin bosons propagate on the null cone. The spin $s$ theory
with helicities $(\pm n,\ldots, 0)$ removed appears when
\be
m^2=M^2\,\Big(s(s-1)-n(n+1)\Big)\, ,
\ee 
and has Bessel index $\nu=n+1/2$. All these gauge theories are unitary.

\vspace{.4cm}

Note that the strictly massless theory has $n=s-1$, $m^2=0$ and $\nu=s-1/2$
whereas the conformal value $\nu=1/2$ 
appears for the scalar gauge invariant theory with
$n=0$ and $m^2=M^2\,s(s-1)$. 
An additional strong justification for this conjecture is that 
these representations have been seen before
in a study~\cite{Higuchi:1987wu} of bosonic Laplace--Beltrami operators
on dS backgrounds\footnote{We note that for $s\geq5/2$,  
massive theories require auxiliary fields; these were 
explicitly seen to decouple
in the $s=5/2,3$ strictly massless limits~\cite{Deser:2001pe}. While we
have no proof that this decoupling persists for higher $s$, the
representations of~\cite{Higuchi:1987wu} suggest this is the case.}.

\section{Null Propagation of Partially Massless Fermions}

Partially massless fermionic theories are found in AdS. However, we continue
to work in dS because of the simplicity of the 
metric~\eqn{dS}. The
results for the partially massless lines depend on $m^2$ and
$\L\equiv3M^2$ 
only and continue to AdS \footnote{The 
cost is that, in dS, the mass parameter $m$ is pure imaginary. The action
is therefore no longer hermitean, although these theories actually possess
a positive norm Hilbert space. This is the higher spin generalization of
the old dS supergravity conundrum discussed in~\cite{Pilch:1985aw}.}. 
Just as for the lowest ``multiline'' $s=5/2$ case, 
only the strictly massless helicity $\pm s$ gauge
theory is unitary in AdS, owing to the line ordering problem discussed 
in~\cite{Deser:2001pe}.   

A massive spin $s\equiv \s+1/2$ fermionic field 
satisfies the onshell conditions
\be
(\Sl D+m)\,\psi_{\m_1\ldots\m_\s}=0\, \qquad D.\psi_{\m_2\ldots\m_\s}=0\
=\g.\psi_{\m_2\ldots\m_\s}\, .
\ee 
We choose the local Lorentz gauge
\be
e_0{}^{\ul 0}=1\, , \qquad e_i{}^{\ul j}=f(t)\,\delta_i{}^{\ul j}\, ,
\ee
where underlined indices are flattened. The Dirac equation for the 
(spatially transverse, gamma-traceless ``$tt$'') helicities $\pm s$
reads
\be
\Big(
\frac{d}{dt}+(2-s)\,M-f^{-1}\,\g^{\ul{0j}}\,\d_j-\g^{\ul 0}\, m
\Big)\,\psi^{tt}_{i_1\ldots i_\s}=0\, .
\ee
In the usual large/small component basis (suppressing indices again)
\be
\g^{\ul 0}=\left(
\begin{array}{rr}
-i&0\\ \;0&i
\end{array}\right)\, ,\qquad
\g^{\ul j}=
\left(
\begin{array}{cc}
0&i\s^j \\ -i\s^j&0
\end{array}\right)\, ,\qquad
\psi^{tt}=\left(
\begin{array}{c}
\chi\\ \phi
\end{array}
\right)\, ,
\ee
we can eliminate the small component $\phi$ and obtain the second
order
equation for $\chi$ 
\be
\Big(
-\frac{d^2}{dt^2}+(2s-5)\,M\,\frac{d}{dt}
+f^{-2}\,\vec\d\,^2-(s-2)(s-3)\,M^2-imM-m^2
\Big)\,\chi=0\, .
\ee
The coordinate transformation~\eqn{change} and field redefinition
\be
\chi\equiv z^{5/2-s}\,q\, ,
\ee
yield Bessel's equation~\eqn{Bessel} with index
\be
\nu^2=\frac{1}{4}-\frac{im}{M}-\frac{m^2}{M^2}=
\Big(\frac{1}{2}-\frac{im}{M}\Big)^2\, .
\ee
Note that $m$ itself will be imaginary for the 
partially massless lines when they are in dS, so the appearance of an 
explicit $im$ here is appropriate.

\subsubsection*{Examples} 

\begin{enumerate}
\item {\em Spin 1/2}: The $m=0$ spin~1/2 theory is well known to be Weyl 
invariant, and indeed we find $\nu=1/2$, the conformal value.
\item {\em Spin 3/2}: As follows from linearizing cosmological supergravity,
spin~3/2 is strictly massless at $m^2=-M^2$. The 
choice of branch $m=iM$, justified by the results, yields $\nu=3/2$.
This is the same value we found above for its strictly massless 
spin~2 superpartners\footnote{We thank K. Peeters for a clarifying 
discussion on this issue.}. 
\item{\em Spin 5/2}: The strictly massless theory is at $m^2=-4M^2$;
the choice $m=2iM$ yields $\nu=5/2$ and null propagation.
The model with a spinor gauge invariance at $m^2=-M^2$ has $\nu=3/2$,
just as for linearized cosmological supergravity.
\end{enumerate}
All the above fermionic partially massless theories propagate on the
null cone\footnote{The argument of~\cite{Deser:1983tm} shows that
spin~3/2 cannot take the conformal value $\nu=1/2$. This 
does {\it not} imply that strictly massless spin~3/2 propagates
off-cone, since null propagation is achieved there by $\nu=3/2$.
This null propagation 
was already proven in~\cite{Deser:1977uq}.}. 
Even though the fermionic analogs of the bosonic 
representations~\cite{Higuchi:1987wu}
have yet to be completed and the above results also depended on a judicious
choice for the branch of the square root $\sqrt{-M^2}$, we make the 
following conjecture:

\vspace{.4cm}

\noindent {\bf Conjecture}: All partially massless fermions propagate on the
null cone. The spin~$s$ theory with helicities $(\pm(n+1/2),\ldots,\pm1/2)$
removed appears at
\be
m^2=-M^2\,(n+1)^2\, 
\ee
with Bessel index $\nu=3/2+n$. The Hilbert space of these theories
is unitary in dS, where their actions are not hermitean. 

\vspace{.4cm}

In particular, the strictly
massless theory with $n=s-3/2$ occurs at $m^2=-M^2\,(s-1/2)^2$
for $s>1/2$: Its Hilbert space is 
unitary for any $\L$, but hermiticity of its action is
lost in dS.
The conformal value $\nu=1/2$ is only attained by the $m=0$ spin~1/2 theory.

\section{Cosmological Speculations}

Having presented an expeditious determination
of the slopes of all partially massless lines, we now present
some speculations based on the results:

\subsection*{The Cosmological Constant Problem}

Previous attempts to render $\L$ small or vanishing
have been of two types: (i)~Those based on (unbroken) symmetries, such as the
cancellation of supersymmetric zero point energies~\cite{Zumino:1975bg} or
the necessary absence of a cosmological terms in $d=11$
supergravity~\cite{Bautier:1997yp}. (ii) A dynamical solution based
on quantum gravity loop corrections driving $\L$ to 
zero~\cite{Tsamis:1993sx}.
Our idea is rather different: it depends only on the kinematics of a 
tower of free fields. Such a leap of faith should not be 
foreign to string theorists, since
massive string states couple an infinite tower of higher spins.

The argument is that the region of the $(m^2,\L)$ plane
where the entire tower has unitary content only is squeezed onto
the $\L=0$ axis:
The unitary region for massive higher spins in the $(m^2,\L)$ plane
is bounded below in AdS by the strictly massless fermionic line
\be
m^2=-\frac{\L\,(s-1/2)^2}{3}\, ,
\ee
and above by the lowest partially massless bosonic
gauge line (the one excluding helicity 0),
\be
m^2=\frac{\L\,s(s-1)}{3}\, .
\ee
Therefore the unitarily allowed region is pinched around $\L=0$ for
large values of $s$. Of course the robustness of this mechanism 
will be challenged by 
the usual array of difficulties introduced by interactions. 

\subsection*{String Theory}

Low energy effective field theories are obtained from strings 
by integrating out
massive string states. Is the existence of new partially massless
theories in cosmological backgrounds relevant to the quantization of 
strings in these spaces?

\subsection*{Supergravities}

A useful confirmation of our calculation was that both strictly massless
spin~3/2 and spin~2 had the same Bessel index, $\nu=3/2$, 
a result that ought be
guaranteed by supersymmetry. Can one find other higher spin
supermultiplets with equal values
of $\nu$ as the basis of new locally supersymmetric theories
in cosmological backgrounds? In the same highly speculative vein, we note 
that the purely imaginary value of $m$ for the dS
partially massless fermionic theories turned out to ensure
real values of the index $\nu$. Could this remark be relevant
to defining a consistent 
cosmological supergravity theory within the dS horizon?

\section*{Acknowledgments}
We thank A. Higuchi, K. Peeters, R. Woodard
and M. Zamaklar for enlightening discussions. 
This work was supported by the 
National Science Foundation under grant PHY99-73935.


\begin{thebibliography}{99} 

\bibitem{Carroll:2000fy}
See for example 
S.~M.~Carroll,
``The cosmological constant,''
astro-ph/0004075.

\bibitem{Maldacena:1998re}
J.~Maldacena,
Adv.\ Theor.\ Math.\ Phys.\  {\bf 2}, 231 (1998)
[Int.\ J.\ Theor.\ Phys.\  {\bf 38}, 1113 (1998)]
[hep-th/9711200]. For a review, see
O.~Aharony, S.~S.~Gubser, J.~Maldacena, H.~Ooguri and Y.~Oz,
Phys.\ Rept.\  {\bf 323}, 183 (2000)
[hep-th/9905111].

\bibitem{Pilch:1985aw}
K.~Pilch, P.~van Nieuwenhuizen and M.~F.~Sohnius,
Commun.\ Math.\ Phys.\ {\bf 98}, 105 (1985).

\bibitem{Deser:2001pe}
S.~Deser and A.~Waldron,
Phys. Rev. Lett., to appear,
[hep-th/0102166];
Nucl. Phys. B, to appear, [hep-th/0103198];
Phys. Lett. B, to appear, [hep-th/0103255].

\bibitem{Deser:1983tm}
S.~Deser and R.~I.~Nepomechie,
Phys.\ Lett.\ B {\bf 132}, 321 (1983);
Annals Phys.\ {\bf 154}, 396 (1984).

\bibitem{Higuchi:1987py}
A.~Higuchi,
Nucl.\ Phys.\ B {\bf 282}, 397 (1987);
{\it ibid} {\bf 325}, 745 (1989);

\bibitem{Higuchi:1987wu}
A.~Higuchi,
J.\ Math.\ Phys.\ {\bf 28}, 1553 (1987).
See also U. Ottoson, Commun. Math. Phys.\ {\bf 8}, 228 (1968);
M. A. Neumark, Am. Math. Soc. Transl. (2)\ {\bf 6}, 337 (1957);
H. Thomas, Am. Math. {\bf 42}, 113 (1941); J. Dixmier, Bull. Soc.
Math. France\ {\bf 89}, 9 (1961); J.G. Kuriyan, N. Mukunda
and E.C.G. Sudarshan, Commun. Math. Phys.\ {\bf 8}, 204 (1968).

\bibitem{Deser:1977uq}
S.~Deser and B.~Zumino,
Phys.\ Rev.\ Lett.\  {\bf 38}, 1433 (1977).

\bibitem{Zumino:1975bg}
B.~Zumino,
Nucl.\ Phys.\ B {\bf 89}, 535 (1975).

\bibitem{Bautier:1997yp}
K.~Bautier, S.~Deser, M.~Henneaux and D.~Seminara,
Phys.\ Lett.\ B {\bf 406}, 49 (1997)
[hep-th/9704131].

\bibitem{Tsamis:1993sx}
N.~C.~Tsamis and R.~P.~Woodard,
Phys.\ Lett.\ B {\bf 301}, 351 (1993);
Annals Phys.\  {\bf 238}, 1 (1995).


\end{thebibliography}
\end{document}